\documentclass[a4paper,11pt]{article}
\pdfoutput=1 

\usepackage{amssymb}
\usepackage{amsmath}
\usepackage{color}
\usepackage{float}
\usepackage[caption = false]{subfig}
\usepackage{jcappub} 

\usepackage[T1]{fontenc} 

\title{\boldmath Stellar equilibrium configurations of compact stars in $f(R,T)$ theory of gravity}
\author{P.H.R.S. Moraes,}
\author{Jos\'e D.V. Arba\~nil,}
\author{M. Malheiro}
\affiliation{ITA - Instituto Tecnol\'ogico de Aeron\'autica - Departamento de F\'isica, 12228-900, S\~ao Jos\'e dos Campos, S\~ao Paulo, Brazil}
\emailAdd{moraes.phrs@gmail.com}
\emailAdd{arbanil@ita.br}
\emailAdd{malheiro@ita.br}
\abstract{In this article we study the hydrostatic equilibrium configuration 
of neutron stars and strange stars, whose fluid pressure is computed from the 
equations of state $p=\omega\rho^{5/3}$ and $p=0.28(\rho-4{\cal B})$, 
respectively, with $\omega$ and ${\cal B}$ being constants and $\rho$ the 
energy density of the fluid. We start by deriving the hydrostatic equilibrium 
equation for the $f(R,T)$ theory of gravity, with $R$ and $T$ standing 
for the Ricci scalar and trace of the energy-momentum tensor, respectively. 
Such an equation is a generalization of the one obtained from general 
relativity, and the latter can be retrieved for a certain limit of the theory. 
For the $f(R,T)=R+2\lambda T$ functional form, with $\lambda$ being a 
constant, we find that some physical properties of the stars, such as 
pressure, energy density, mass and radius, are affected when $\lambda$ is 
changed. We show that for a fixed central star energy density, the mass of neutron and strange stars can increase with $\lambda$. Concerning the star radius, it increases for neutron stars and it decreases for strange stars with the increment of $\lambda$. Thus, in $f(R,T)$ theory of gravity we can push the maximum mass above the observational limits. This implies that the equation of state cannot be eliminated if the maximum mass within General Relativity lies below the limit given by observed pulsars. 
}

\begin{document}
\maketitle
\flushbottom

\section{Introduction}\label{sec:int}

Modified gravity theories have been constantly proposed with the purpose of solving (or evading) the $\Lambda$CDM cosmological model shortcomings (check \cite{clifton/2012} for instance). 

Anyhow, $\Lambda$CDM model indeed provides a great matching between theory predictions and observational data. However, the fundamental physical properties of the matter field responsible for the observationally predicted cosmic acceleration \cite{riess/1998,perlmutter/1999} in the so-called ``concordance model" still lacks a convincing explanation.

Nevertheless it is remarkable the fact that a good agreement between theory and observation may be obtained also from cosmological models derived from modified gravity theories (check \cite{garcia-bellido/2008,bahcall/1999}, for instance).

It should be mentioned here that a gravity theory must be tested also in the astrophysical level \cite{astashenok/2013,staykov/2014,astashenok/2015b,astashenok/2015}. Strong gravitational fields found in relativistic stars could discriminate standard gravity from its generalizations. Observational data on neutron stars (NSs), for instance, can be used to investigate possible deviations from General Relativity (GR) as probes for modified gravity theories.

NSs are the best natural laboratories to test gravity in the strong field regime. Their collapse leads to large-curvature and strong-gravity environments \cite{psaltis/2008}. Macroscopic properties of NSs are sensitive to the equation of state (EoS) and the underlying gravitational theory.

The modified theory to be investigated in this article will be the $f(R,T)$ gravity \cite{harko/2011}. Originally proposed as a generalization of the $f(R)$ gravity \cite{nojiri/2007,nojiri/2009}, the $f(R,T)$ theories assume that the gravitational part of the action still depends on a generic function of the Ricci scalar $R$, but also presents a generic dependence on $T$, the trace of the energy-momentum tensor. Such a dependence on $T$ would come from the consideration of quantum effects.

Although a lot of cosmological applications to $f(R,T)$ gravity have already been proposed (check \cite{ms/2016,moraes/2015,moraes/2015b,moraes/2014b,farasat_shamir/2015,singh/2014,shabani/2014,shabani/2013} and references therein), at an astrophysical level, the theory still lacks a deep research. In \cite{sharif/2014}, the authors have investigated the stability of collapsing spherical body coupled with isotropic matter from a non-static spherically symmetric line element. In \cite{noureen/2015}, it was employed the perturbation scheme to find the collapse equation. Conditions on the adiabatic index were constructed for Newtonian and post-Newtonian eras in order to address the  instability problem. In \cite{noureen/2015b}, the authors have developed the instability range of the $f(R,T)$ theory for an anisotropic background constrained by zero expansions. The evolution of a spherical star by employing a perturbation scheme on the $f(R,T)$ field equations was explored in \cite{noureen/2015c}, while the dynamical analysis for gravitating sources carrying axial symmetry was discussed in \cite{zubair/2015}. Moreover, very recently, in \cite{zubair/2016}, the formation of spherically symmetric anisotropic stars was investigated in the context of $f(R,T)$ theory. 

Although the dynamics of collapse and the instability range have indeed been investigated in $f(R,T)$ gravity, a self-consistent model of compact star, with a specific EoS, obtained from the solution of the Tolman-Oppenheimer-Volkoff (TOV) equation \cite{tolman,oppievolkoff} has not been derived so far for such a theory.

As mentioned above, not only the EoS plays an important role in predicting macroscopic properties of compact objects. The underlying gravity theory is also fundamental on this regard. For instance, in the Randall-Sundrum type II brane model \cite{randall/1999}, it was found that the hadronic and strange stars (SSs) can violate the GR causal limit for large enough masses \cite{arbanil_lugones_proc,arbanil_lugones_article,castro2014}.

Moreover, the TOV equations derived for different modified gravity models \cite{astashenok/2015,astashenok/2015b,momeni/2015,momeni/2015b} have contributed to the understanding of compact stars structure and matter at high densities. The exact conditions under which the fundamental degrees of freedom of strongly interacting matter can be realized in nature is still an open question. A deeper and systematic study of systems containing compact stars may improve our understanding of such an important issue.

In this article, we will investigate the spherical equilibrium configuration of NSs and SSs in $f(R,T)$ theory of gravity.  Here, it is worth mentioning that SSs existence is still a possibility. In fact, a recent work has
discussed the existence of these stars from a gravitational wave detection perspective \cite{moraes/2014c}.
Once such objects existence is proved, the fundamental state of matter at high densities will be understood as the strange quark matter (SQM) state.

This article is structured as follows. In Sec.~\ref{sec:frt} we briefly review the $f(R,T)$ theory of gravity. In Sec.~\ref{sec:tovfrt} we present the stellar structure equations, the boundary conditions and the EoS used to analyze the NSs and SSs in $f(R,T)$ gravity. In Sec.~\ref{sec:sol} the equilibrium configurations of NSs and SSs in $f(R,T)$ gravity are presented. In Sec.~\ref{sec:dis} we present our discussion and conclusions. Throughout this article we will work with the geometrized units system, i.e., $c=1=G$.

\section{The $f(R,T)$ theory of gravity}\label{sec:frt}

Proposed by T. Harko et al., the $f(R,T)$ gravity \cite{harko/2011} assumes the gravitational part of the action depends on a generic function of $R$ and $T$, the Ricci scalar and the trace of the energy-momentum tensor $T_{\mu\nu}$, respectively. By assuming a matter lagrangian density $\mathcal{L}_m$, the total action reads
\begin{equation}\label{frt1}
S=\frac{1}{16\pi}\int d^{4}xf(R,T)\sqrt{-g}+\int d^{4}x\mathcal{L}_m\sqrt{-g}.
\end{equation} 
In (\ref{frt1}), $f(R,T)$ is the generic function of $R$ and $T$, and $g$ is the determinant of the metric tensor $g_{\mu\nu}$.

By varying (\ref{frt1}) with respect to the metric $g_{\mu\nu}$, one obtains the following field equations:
\begin{eqnarray}\label{frt2}
\hspace{-0.2cm}f_R(R,T)R_{\mu\nu}-\frac{1}{2}f(R,T)g_{\mu\nu}+(g_{\mu\nu}\Box-\nabla_\mu\nabla_\nu)\nonumber\\
\hspace{-0.2cm}f_R(R,T)=8\pi T_{\mu\nu}-f_T(R,T)T_{\mu\nu}-f_T(R,T)\Theta_{\mu\nu},
\end{eqnarray}
in which $f_R(R,T)\equiv\partial f(R,T)/\partial R$, $f_T(R,T)\equiv\partial f(R,T)/\partial T$, $\Box\equiv\partial_\mu(\sqrt{-g}g^{\mu\nu}\partial_\nu)/\sqrt{-g}$, $R_{\mu\nu}$ represents the Ricci tensor, $\nabla_\mu$ the covariant derivative with respect to the symmetric connection associated to $g_{\mu\nu}$, $\Theta_{\mu\nu}\equiv g^{\alpha\beta}\delta T_{\alpha\beta}/\delta g^{\mu\nu}$ and $T_{\mu\nu}=g_{\mu\nu}\mathcal{L}_m-2\partial\mathcal{L}_m/\partial g^{\mu\nu}$.

Taking into account the covariant divergence of (\ref{frt2}) yields \cite{alvarenga/2013,barrientos/2014}
\begin{eqnarray}\label{frt3}
\hspace{-0.5cm}\nabla^{\mu}T_{\mu\nu}&=&\frac{f_T(R,T)}{8\pi -f_T(R,T)}[(T_{\mu\nu}+\Theta_{\mu\nu})\nabla^{\mu}\ln f_T(R,T)\nonumber \\
&&+\nabla^{\mu}\Theta_{\mu\nu}-(1/2)g_{\mu\nu}\nabla^{\mu}T].
\end{eqnarray}

We will assume the energy-momentum tensor of a perfect fluid, i.e., $T_{\mu\nu}=(\rho+p)u_\mu u_\nu-pg_{\mu\nu}$, with $\rho$ and $p$ respectively representing the energy density and pressure of the fluid and $u_\mu$ being the four-velocity tensor, which satisfies the conditions $u_\mu u^{\mu}=1$ and $u^\mu\nabla_\nu u_\mu=0$. We have, then, $\mathcal{L}_m=-p$ and $\Theta_{\mu\nu}=-2T_{\mu\nu}-pg_{\mu\nu}$.

For the functional form of the $f(R,T)$ function above, note that originally suggested by T. Harko et al. in \cite{harko/2011}, the form $f(R,T)=R+2\lambda T$, with $\lambda$ being a constant, has been extensively used to obtain $f(R,T)$ cosmological solutions (check \cite{moraes/2014b,moraes/2015,moraes/2015b,farasat_shamir/2015} and references therein). The $f(R,T)$ gravity authors themselves have derived in \cite{harko/2011} a scale factor which describes an accelerated expansion from such an $f(R,T)$ form. Here, we propose $f(R,T)=R+2\lambda T$ to derive the $f(R,T)$ TOV equation. 

The substitution of $f(R,T)=R+2\lambda T$ in Eq.(\ref{frt2}) yields \cite{moraes/2014b,moraes/2015}
\begin{equation}\label{frt4}
G_{\mu\nu}=8\pi T_{\mu\nu}+\lambda Tg_{\mu\nu}+2\lambda(T_{\mu\nu}+pg_{\mu\nu}),
\end{equation}
for which $G_{\mu\nu}$ is the usual Einstein tensor. 

Moreover, when $f(R,T)=R+2\lambda T$, Eq.(\ref{frt3}) reads
\begin{align}\label{frt5}
\nabla^{\mu}T_{\mu\nu}=-\frac{2\lambda}{8\pi+2\lambda}\left[\nabla^{\mu}(pg_{\mu\nu})+\frac{1}{2}g_{\mu\nu}\nabla^{\mu}T\right].
\end{align}

\section{Equations of stellar structure in $f(R,T)$ gravity}\label{sec:tovfrt}

\subsection{Hydrostatic equilibrium equation}

In order to construct the $f(R,T)$ hydrostatic equilibrium equation, we must, firstly, develop the field equations (\ref{frt4}) for a spherically symmetric metric, such as
\begin{equation}\label{tovfrt1}
ds^2=e^{\vartheta(r)}dt^2-e^{\varpi(r)}dr^2-r^2(d\theta^2+\sin^2\theta d\phi^2).
\end{equation}

For (\ref{tovfrt1}), the non-null components of the Einstein tensor read

\begin{equation}\label{tovfrt2}
G_0^{0}=\frac{e^{-\varpi}}{r^{2}}(-1+e^{\varpi}+\varpi' r),
\end{equation}
\begin{equation}\label{tovfrt3}
G_1^{1}=\frac{e^{-\varpi}}{r^{2}}(-1+e^{\varpi}-\vartheta' r),
\end{equation}
\begin{equation}\label{tovfrt4}
G_2^{2}=\frac{e^{-\varpi}}{4r}[2(\varpi'-\vartheta')-(2\vartheta''+\vartheta'^{2}-\vartheta'\varpi')r],
\end{equation}
\begin{equation}\label{tovfrt5}
G_3^{3}=G_2^{2},
\end{equation}
for which primes stand for derivations with respect to $r$.

By substituting Eqs.(\ref{tovfrt2})-(\ref{tovfrt3}) in (\ref{frt4}) yields

\begin{equation}\label{tovfrt6}
\frac{e^{-\varpi}}{r^{2}}(-1+e^{\varpi}+\varpi'r)=8\pi\rho+\lambda(3\rho-p),
\end{equation}
\begin{equation}\label{tovfrt7}
\frac{e^{-\varpi}}{r^{2}}(-1+e^{\varpi}-\vartheta'r)=-8\pi p+\lambda(\rho-3p).
\end{equation}

As usually, we introduce the quantity $m$, representing the gravitational mass within the sphere of radius $r$, such that $e^{-\varpi}=1-2m/r$. Replacing it in \eqref{tovfrt6} yields

\begin{equation}\label{mass_continuity}
m'=4\pi r^2\rho+\frac{\lambda(3\rho-p)r^2}{2}.
\end{equation}

Moreover, from the equation for the non-conservation of the energy-momentum tensor (\ref{frt5}), we obtain

\begin{equation}\label{tovfrt8}
p'+(p+\rho)\frac{\vartheta'}{2}=\frac{\lambda}{8\pi+2\lambda}(p'-\rho').
\end{equation}
Note that in (\ref{frt4}), (\ref{frt5}), (\ref{tovfrt6}), (\ref{tovfrt7}), (\ref{mass_continuity}) and (\ref{tovfrt8}), when $\lambda=0$ the GR predictions are retrieved.

Replacing Eq.\eqref{tovfrt7} in \eqref{tovfrt8} yields a novel hydrostatic equilibrium equation:

\begin{equation}\label{tov}
p'=-(p+\rho)\frac{\left[4\pi pr+\frac{m}{r^2}-\frac{\lambda(\rho-3p)r}{2}\right]}{\left(1-\frac{2m}{r}\right)\left[1-\frac{\lambda}{8\pi+2\lambda}\left(1-\frac{d\rho}{dp}\right)\right]}.
\end{equation}
Note that by taking $\lambda=0$ in Eq.\eqref{tov} yields the standard TOV equation \cite{tolman,oppievolkoff}. We observe that the hydrostatic equilibrium configurations are obtained only when $\frac{\lambda}{8\pi+2\lambda}\left(1-\frac{d\rho}{dp}\right)<1$. It is important to mention that to derive Eq.\eqref{tov}, we considered that the energy density depends on the pressure ($\rho=\rho(p)$).

\subsection{Boundary conditions}

The integration of equations \eqref{mass_continuity} and \eqref{tov} starts with the values in the center ($r=0$): 

\begin{equation}\label{boundary1}
m(0)=0,\,\,\,\,\rho(0)=\rho_{c},\,\,\,\,p(0)=p_{c}.
\end{equation}

The surface of the star ($r=R$) is determined when $p(R)=0$. At the surface, the interior solution connects softly with the Schwarzschild vacuum solution. The potential metrics of the interior and of the exterior line element are linked by $e^{\vartheta(R)}=1/e^{\varpi(R)}=1-2M/R$, with $M$ representing the stellar total mass.

\subsection{Equations of state}

In order to close the system of equations above we will assume a relation between $\rho$ and $p$, known as EoS.
Once defined the EoS, the coupled differential equations \eqref{mass_continuity} and \eqref{tov} can be solved for three unknown functions $m$, $p$ and $\rho$. Recall that these coupled differential equations are integrated from the center towards the surface of the object.

To analyze the equilibrium configurations of compact stars in $f(R,T)$ theory of gravity, two EoS frequently used in the literature will be considered: the polytropic and the MIT bag model EoS.

Within the simplest choices, we find that the polytropic EoS is one of the most used for the study of compact stars. Following the work developed by R.F. Tooper \cite{tooper1964}, we consider that $p=\omega\rho^{5/3}$, with $\omega$ being a constant. We choose the value of $\omega$ to be $1.475\times10^{-3}\,[\rm fm^3/MeV]^{2/3}$ as in \cite{raymalheirolemoszanchin,alz-poli-qbh}.  

To describe SQM, the MIT bag model will be considered. Such an EoS describes a fluid composed by up, down and strange quarks only \cite{witten1984}. It has been applied to investigate the stellar structure of compact stars, e.g., see \cite{farhi_jaffe1984,Malheiro2003}. It is given by the relation $p=a\,(\rho-4{\cal B})$. The constant $a$ is equal to $1/3$ for massless strange quarks and equal to $0.28$ for massive strange quarks, with $m_s=250\,[\rm MeV]$ \cite{stergioulas2003}. The parameter ${\cal B}$ is the bag constant. In this work, we consider $a=0.28$ and ${\cal B}=60\,[\rm MeV/fm^3]$.

\section{Neutron stars and strange stars in $f(R,T)$ gravity}\label{sec:sol}

\subsection{About the numerical method}

After defining the EoS, the stellar structure equations \eqref{mass_continuity} and \eqref{tov} will be solved numerically together with the boundary conditions for different values of $\rho_c$ and $\lambda$, through the Runge-Kutta $4$th-order method. 

\subsection{Energy density, pressure and mass profile in the interior of a star}

\begin{figure*}[ht]
\centering
\subfloat{\includegraphics[scale=0.173]{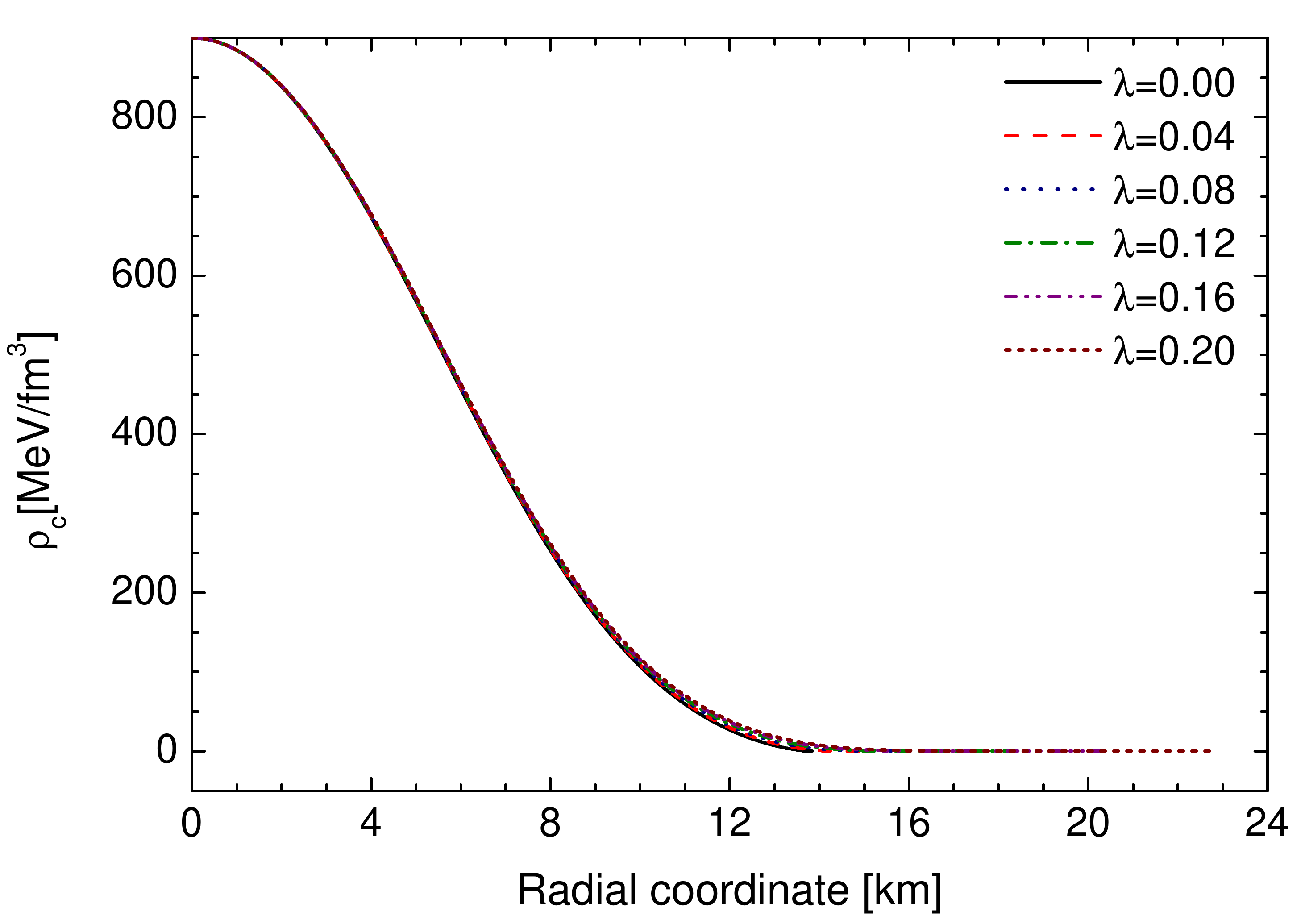}} 
\subfloat{\includegraphics[scale=0.173]{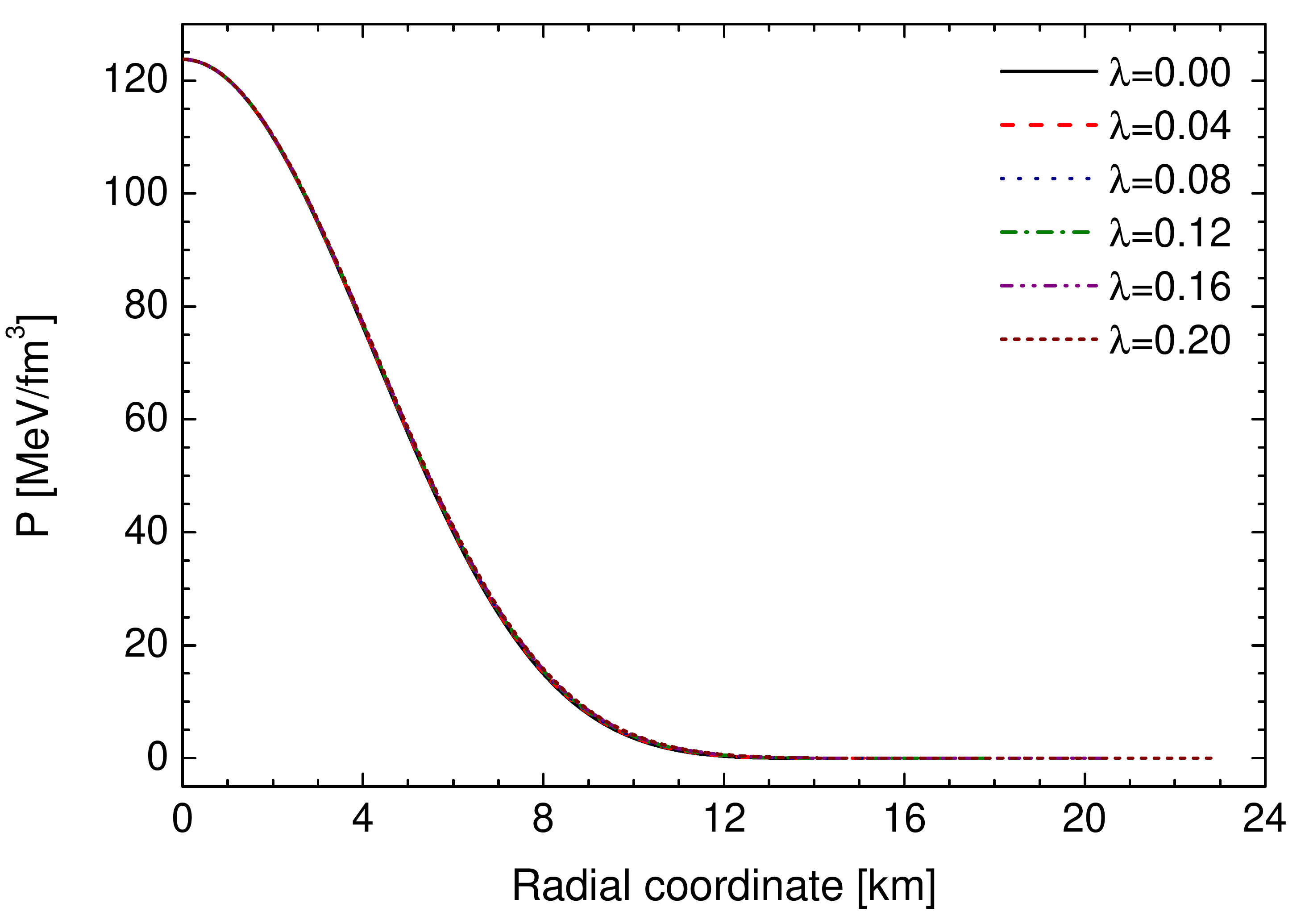}}
\subfloat{\includegraphics[scale=0.173]{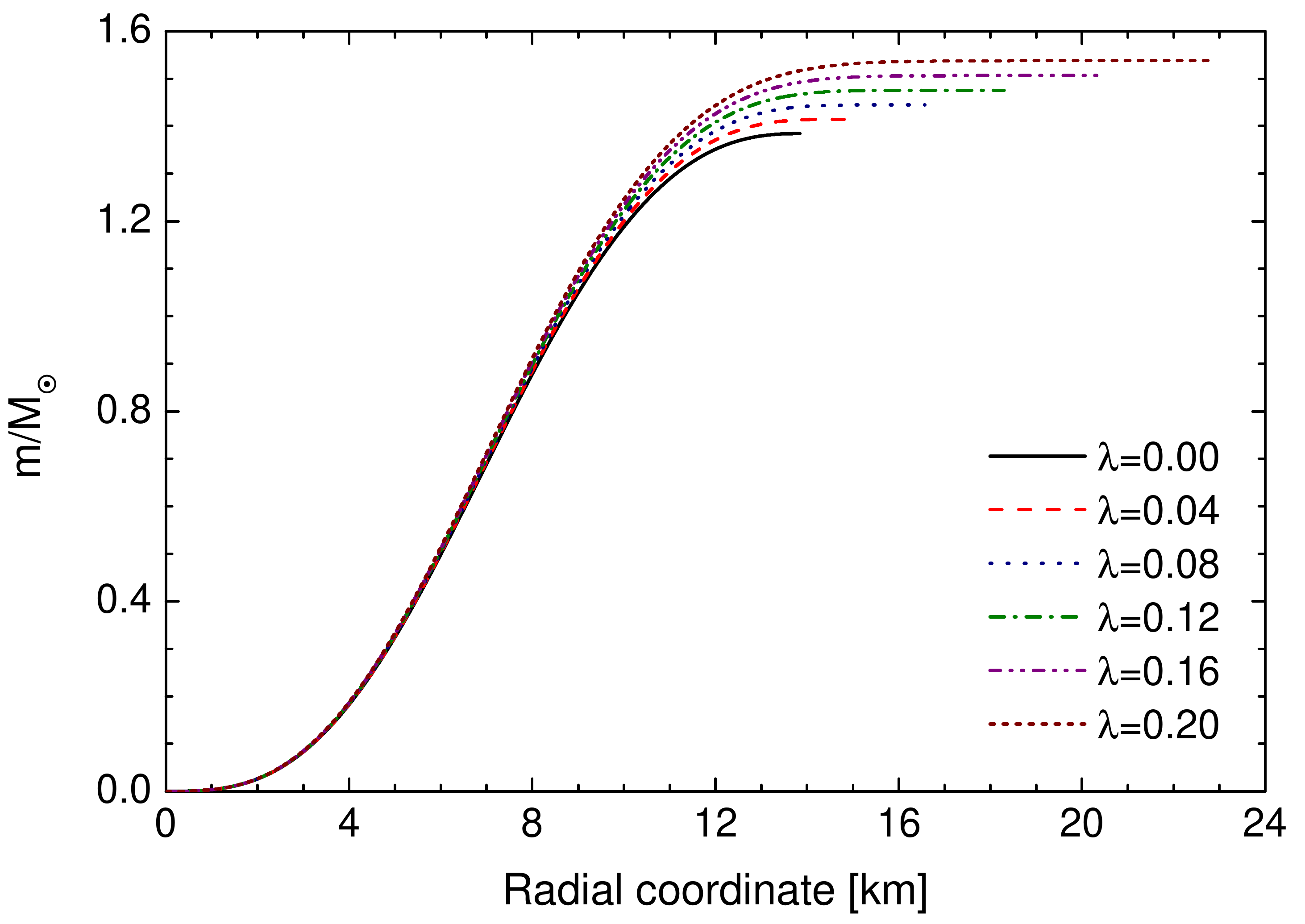}}
\    
\subfloat{\includegraphics[scale=0.173]{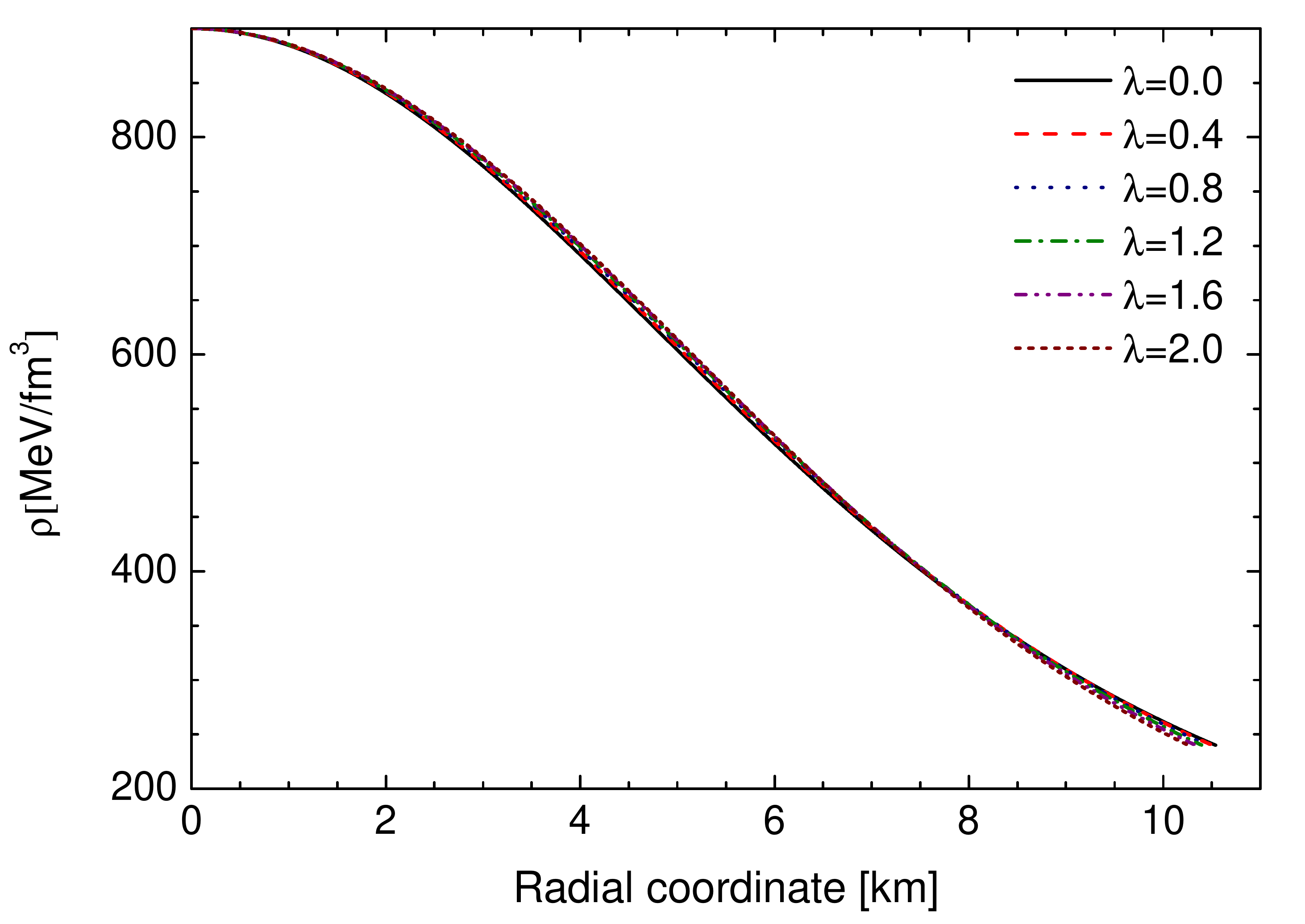}} 
\subfloat{\includegraphics[scale=0.173]{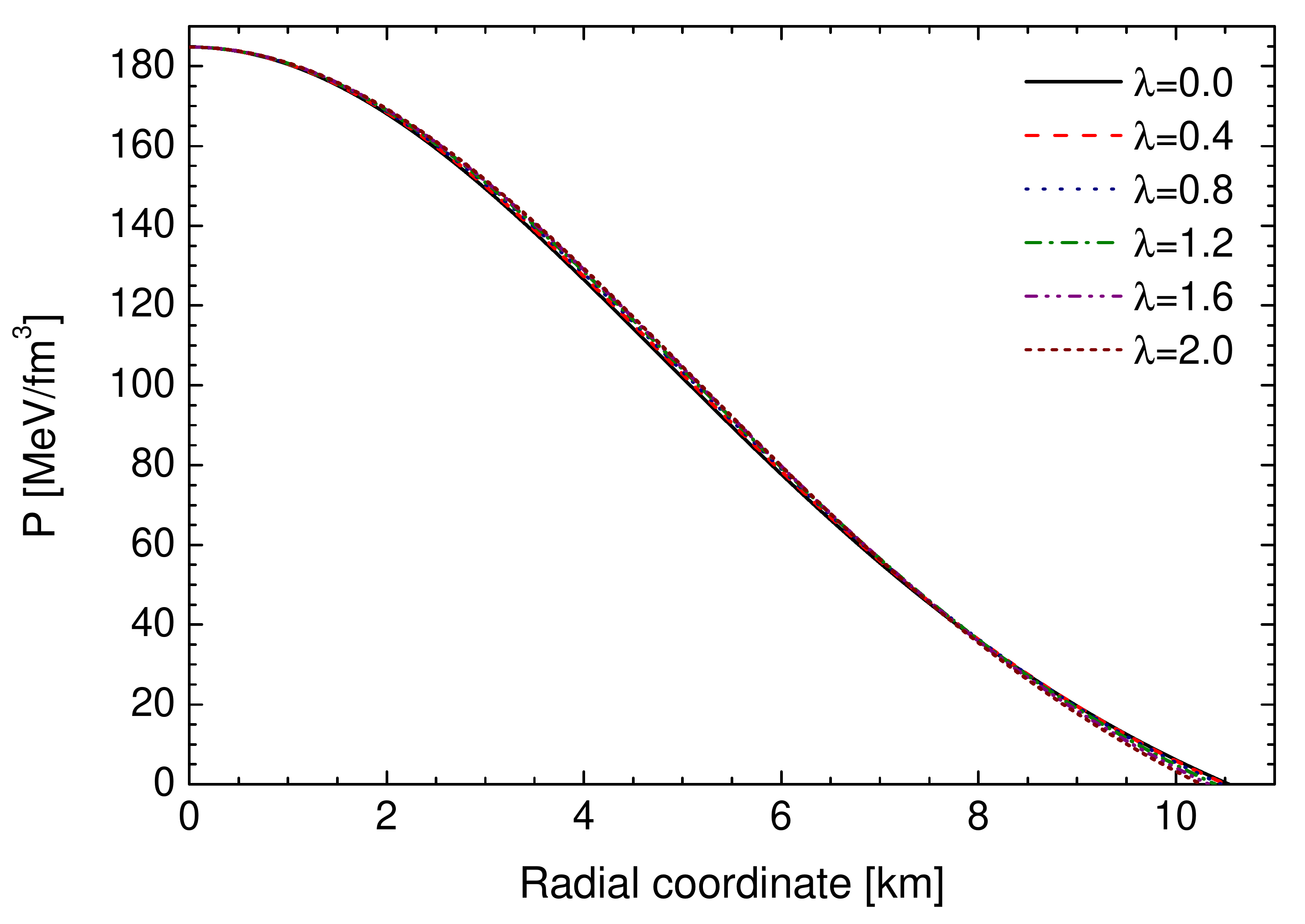}}
\subfloat{\includegraphics[scale=0.173]{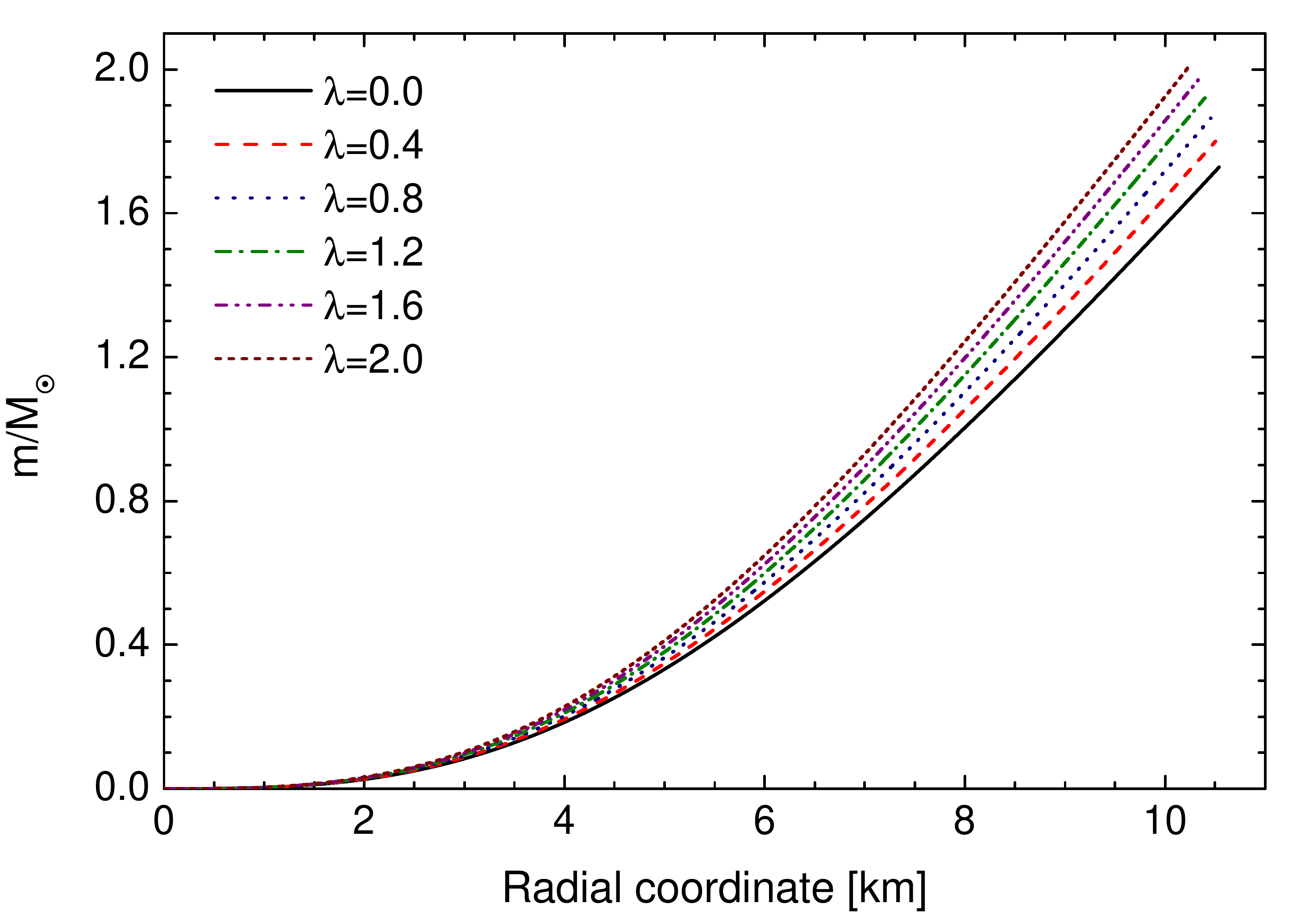}}
\caption{The energy density, pressure and mass (in solar masses) against radial coordinate for neutron and strange stars are presented respectively on the top and bottom panels, for some different values of the parameter $\lambda$. In all cases, it is considered the central energy density $900\,[\rm MeV/fm^3]$.}
\label{rho_p_r}
\end{figure*}

The behavior of the energy density $\rho$, the radial pressure $p$ and the mass $m/M_{\odot}$ as a function of the radial coordinate are presented in Fig.\ref{rho_p_r} for different values of $\lambda$. The functions $\rho(r)$, $p(r)$ and $m(r)/M_{\odot}$ in the figure are constructed using two different EoS and the central energy density $900\,[\rm MeV/fm^3]$. On the top and on the bottom of the figure are depicted the results found for a NS and a SS, respectively, for the GR case ($\lambda=0$) and for the $f(R,T)$ theory of gravity case ($\lambda\neq0$). 

It can be observed an increment of the mass of the stars when $\lambda$ is increased, namely, when the term $2\lambda T$ is incremented. We can, then, say that the effect caused by the term $2\lambda T$ is similar to that caused by an extra pressure or electric charge in the configurations of NSs and SSs in GR scope (see, for instance, \cite{raymalheirolemoszanchin,alz-poli-qbh,arbanil_malheiro2015_2,arbanil_malheiro,negreiros2009}). On the other hand, the total radius of NSs grows with the increment of $\lambda$, in turn, the total radius of SSs decreases with the increase of $\lambda$. The increasing and decreasing of the total radius of NSs and SSs with $\lambda$ are explained in more details in Subsection \ref{Rxrho_fRT}.

\subsection{Equilibrium configurations of neutron stars and strange stars}\label{ss:ns}

\begin{figure}[ht]
\centering
{\includegraphics[scale=0.25]{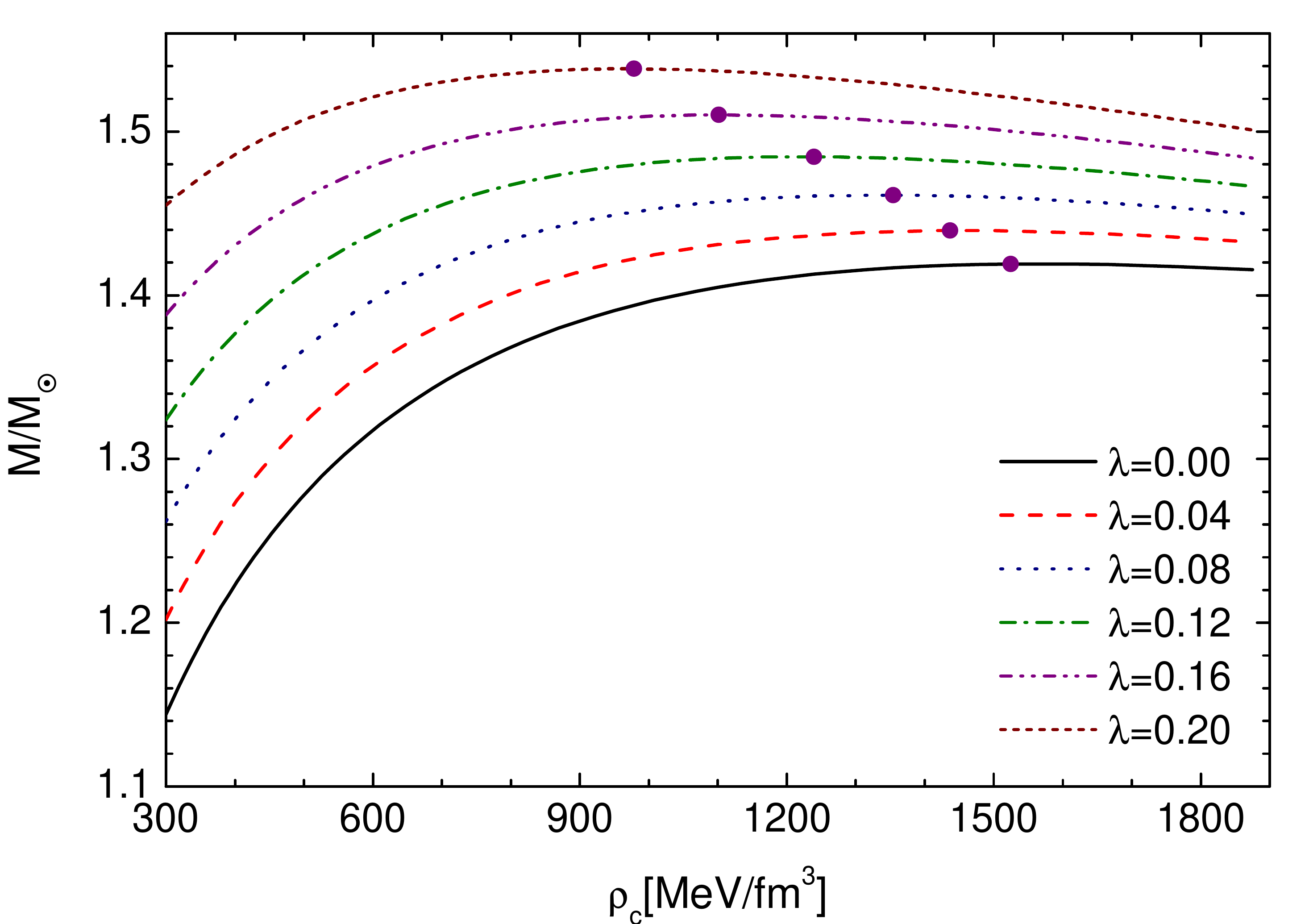}}
{\includegraphics[scale=0.25]{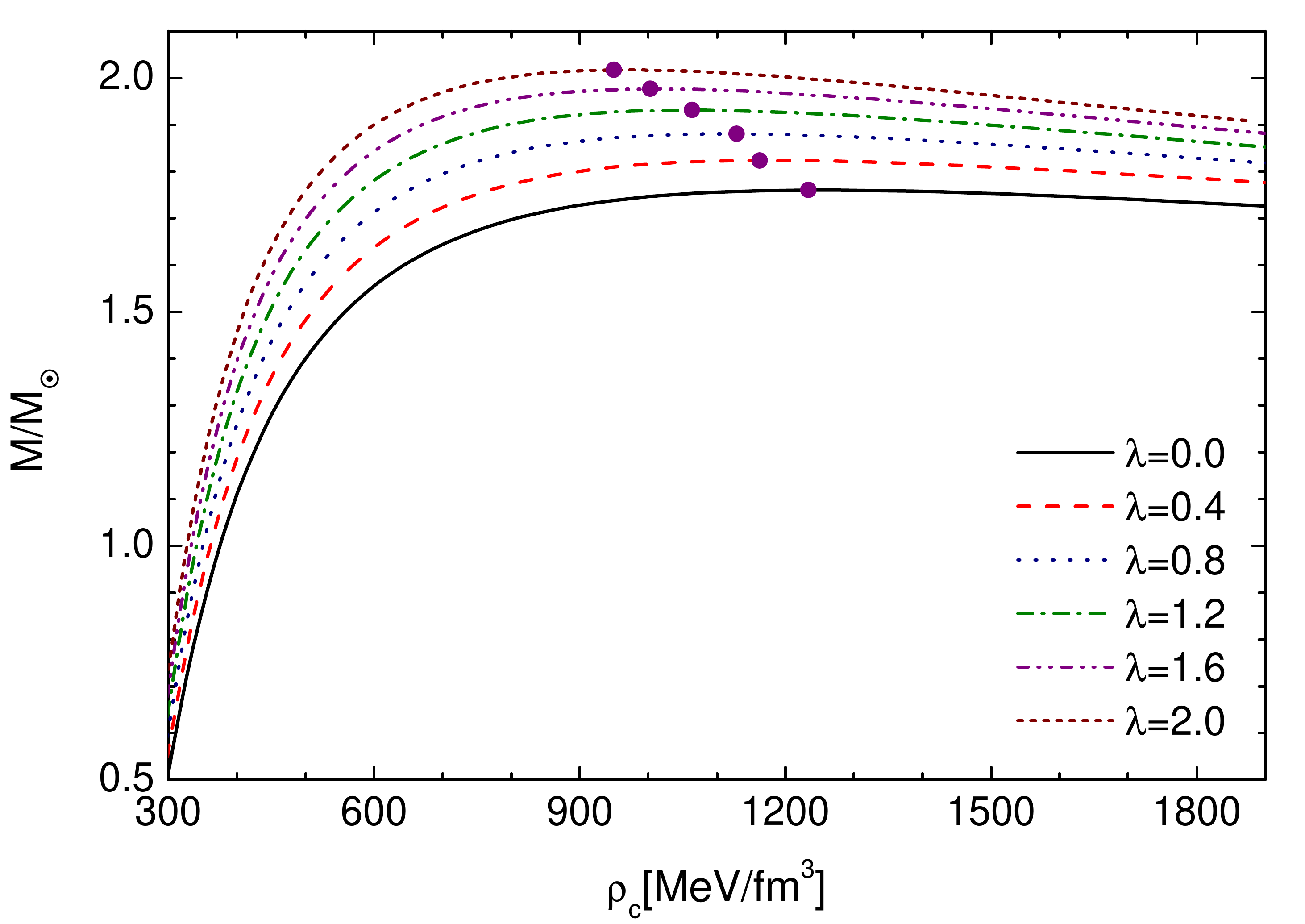}}
\caption{The total mass of the star as a function of the central energy density for different values of $\lambda$. The curves plotted on the left and right panels are characteristics of neutron and strange stars, respectively. The full circles over the curves indicate the maximum mass points.
}
\label{RhoxM_lam}
\end{figure}

The behavior of the total mass, normalized in solar masses, against the central energy density $\rho_c$, for NSs and SSs, is plotted respectively on the left and right panels of Fig.\ref{RhoxM_lam} for different values of the parameter $\lambda$. For NSs and SSs, the masses increase with the central energy density until it reaches its maximum value (marked by a full circle). After this point the mass decreases monotonically with the growth of $\rho_c$.

Note that for the greater values of $\lambda$, the maximum mass point is attained in lower central energy densities. For $\lambda=0$, in the NS case, we determine that the maximum mass point $1.419\,M_{\odot}$ is found for the central energy density $\rho_c\approx10.97\,\rho_{\rm nuclear}$. We also found that when we increment the value of $\lambda$, the maximum mass value also grows. For instance, for $\lambda=0.20$ the maximum mass point $1.538\,M_{\odot}$ is determined when $\rho_c\approx6.969\,\rho_{\rm nuclear}$. In turn, for SSs we determine when $\lambda=0$ that the maximum mass point, $1.760\,M_{\odot}$, is found for $8.917\,\rho_{\rm nuclear}$. On the other hand, for $\lambda=2.0$ we find that the maximum mass $2.017\,M_{\odot}$ is reached for $6.836\,\rho_{\rm nuclear}$.

From this we understand that for a more adequate EoS to study NSs and for a particular value of $\lambda$, the high values of mass observed for the pulsars PSR J$1614-2230$ and PSR J$0348+0432$ \cite{demorest_nature,antoniadis_science}, which still require a convincing explanation, can be easily achieved.


\begin{figure}[ht]
\centering
{\includegraphics[scale=0.25]{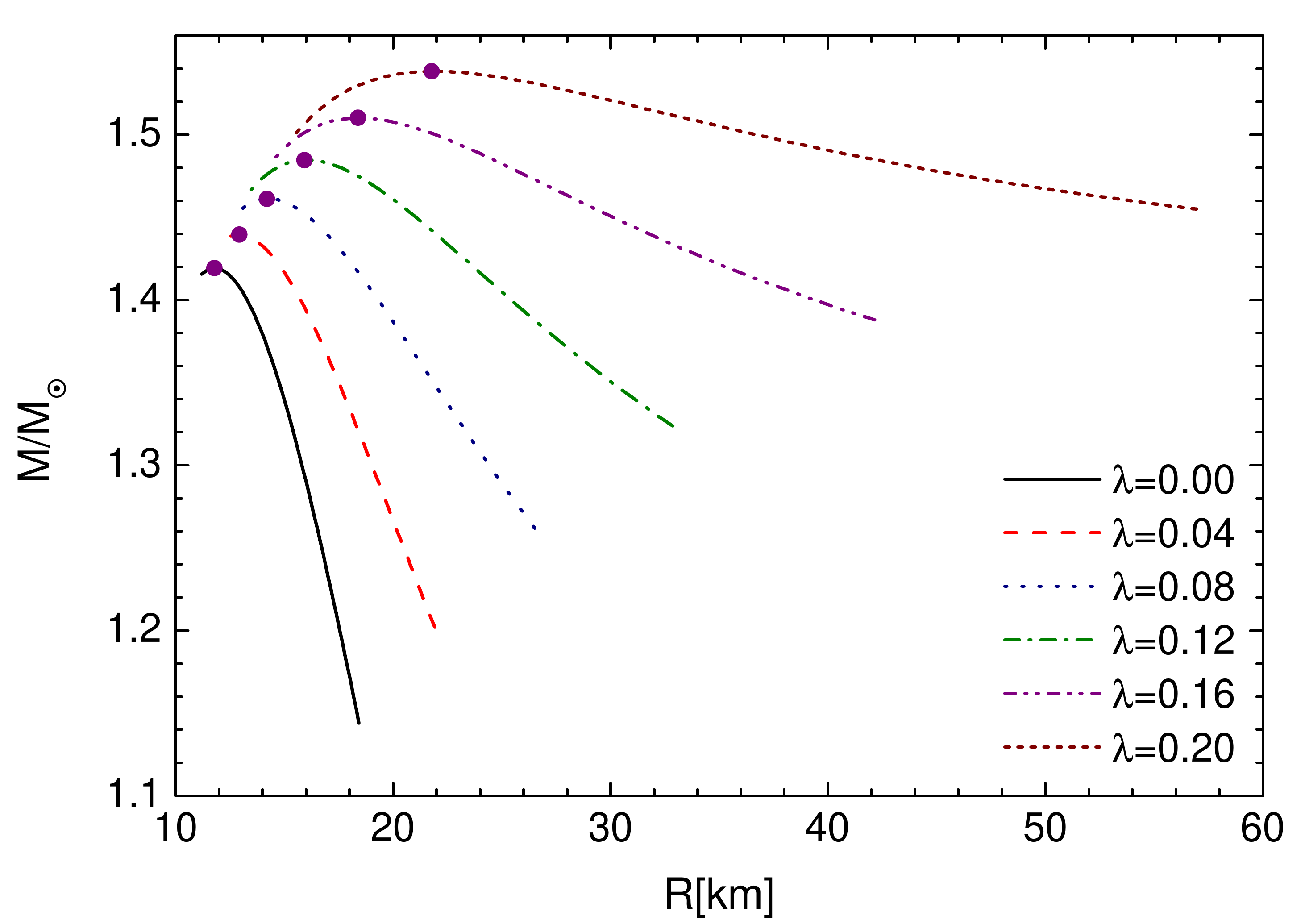}}
{\includegraphics[scale=0.25]{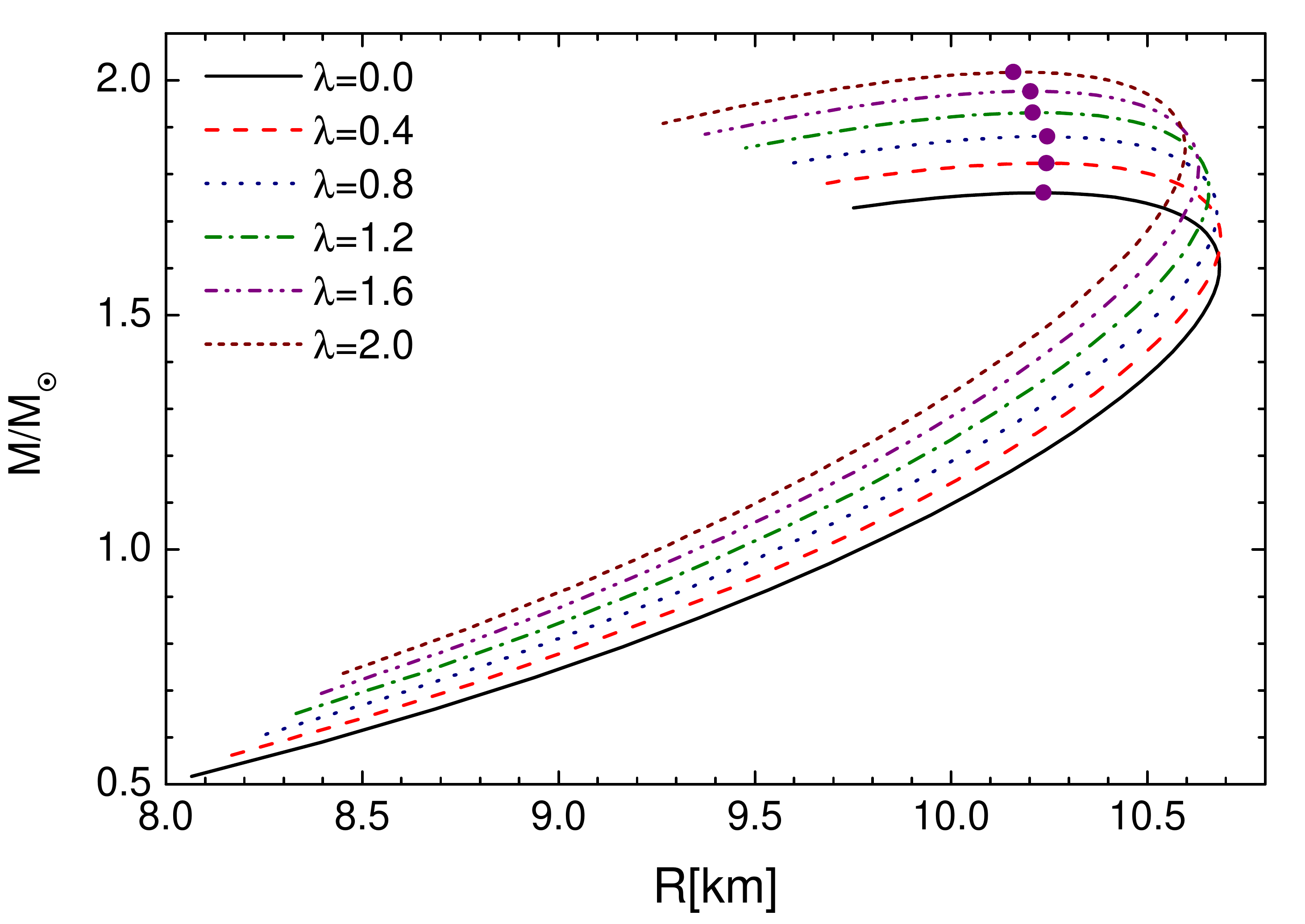}}
\caption{The total mass $M/M_{\odot}$ against the radius of the star for neutron (left) and strange stars (right), for different values of $\lambda$. The maximum mass points on the curves are indicated by full circles.
}
\label{RxM_lam}
\end{figure}

Fig.\ref{RxM_lam} shows the behavior of the total mass, normalized in solar masses, versus the radius of the star, 
for different different values of the parameter $\lambda$. On the left and on the right of Fig.\ref{RxM_lam} are shown 
the results obtained for NSs and SSs, respectively. We can see that when we increase the value of $\lambda$, the NSs become larger and more massive, in turn, the SSs become lower and more massive. It is attributed  to the EoS used in each case . 

\subsection{Total radius of neutron stars and strange stars in $f(R,T)$ gravity}\label{Rxrho_fRT}

\begin{figure}[ht]
\centering
{\includegraphics[scale=0.25]{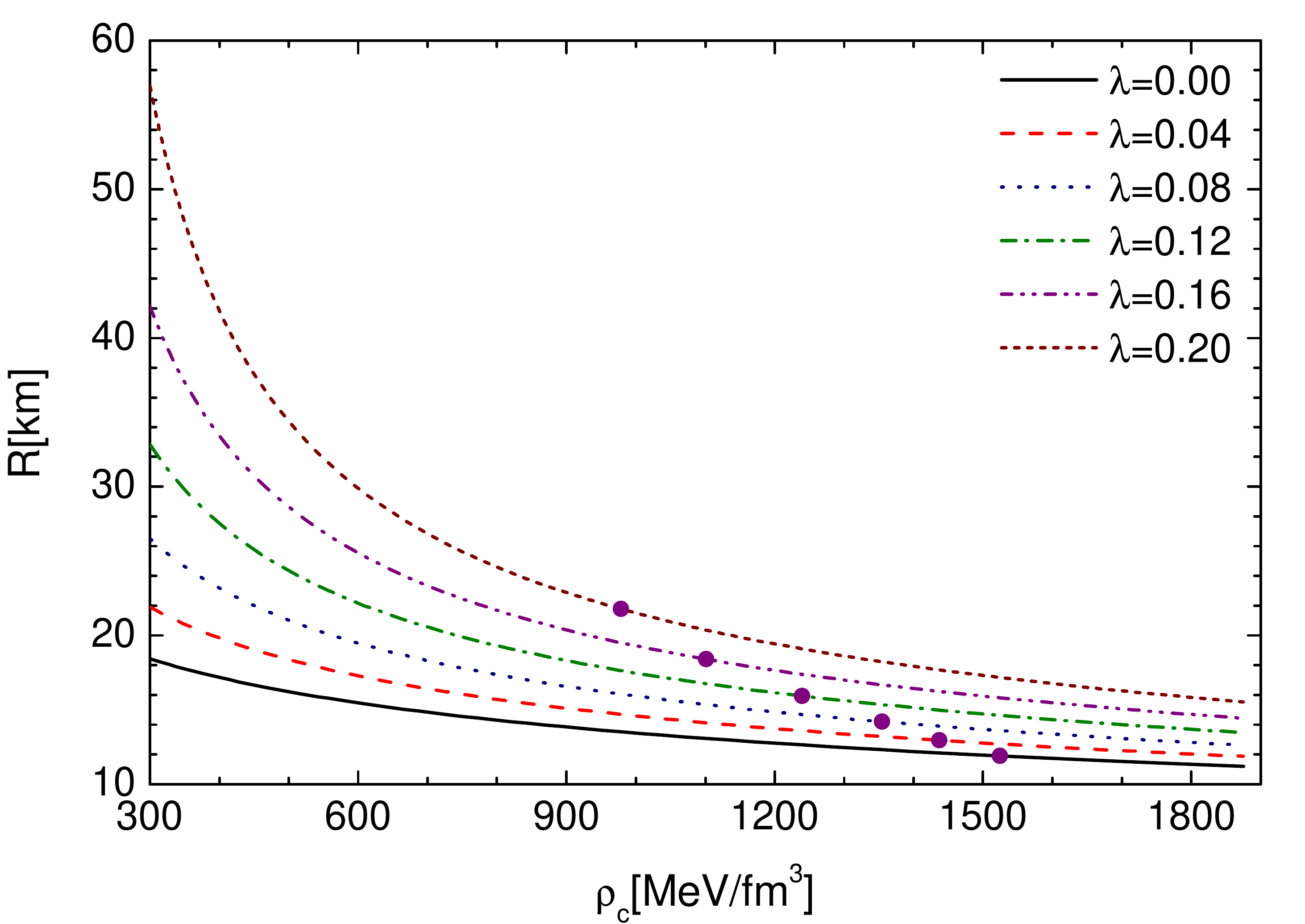}}
{\includegraphics[scale=0.25]{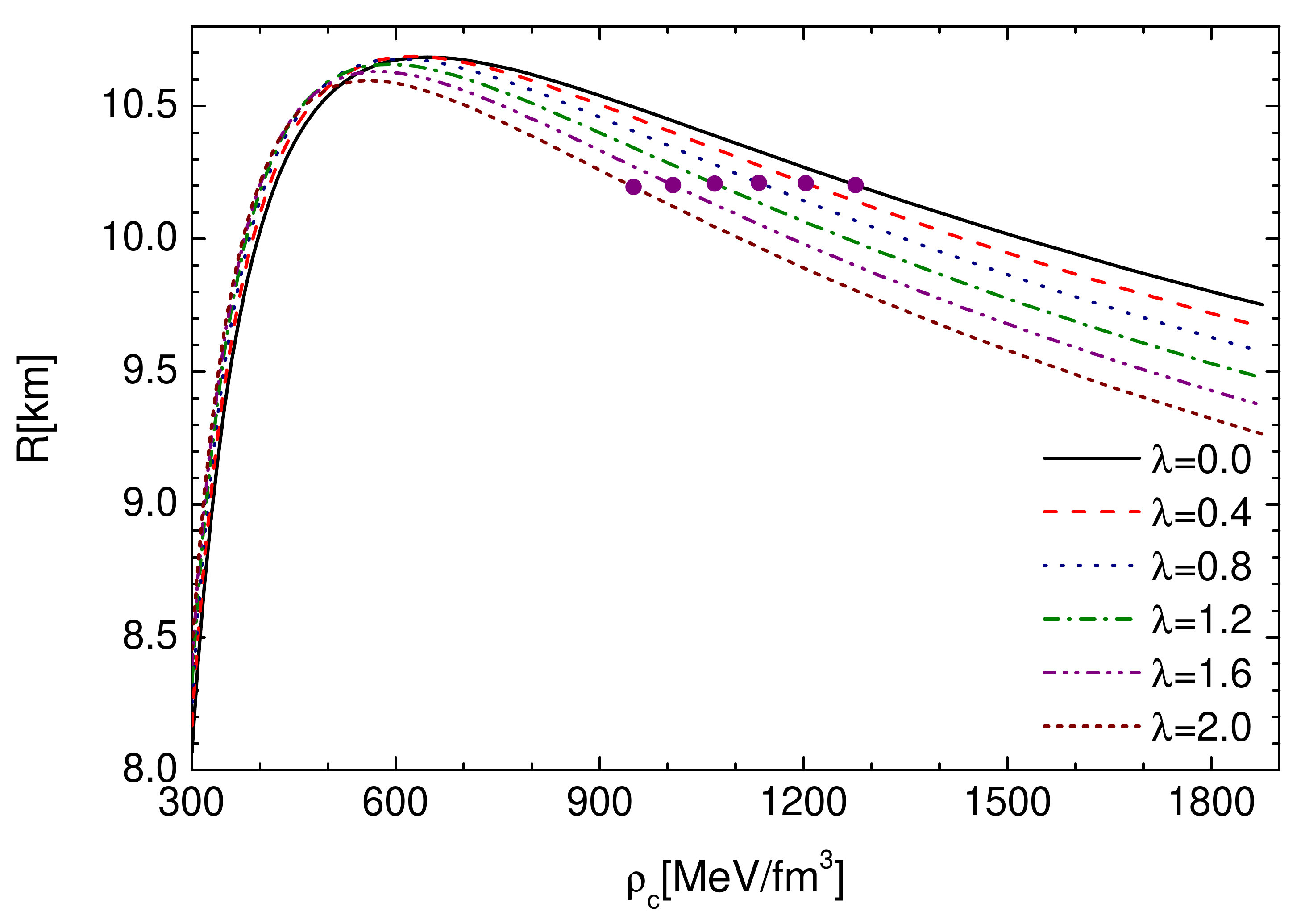}}
\caption{The total radius of the star against the central energy density for some values of $\lambda$. The curves for neutron and strange stars are plotted respectively on the left and right panels. The full circles over the curves represent the places where the maximum mass points are found.
}
\label{rho_R_lam}
\end{figure}

In Fig.\ref{rho_R_lam}, the total radius of NSs and SSs as a function of the central energy density is presented on the left and right panels, respectively. 

Note that for NSs the total radius increases and for SSs it decreases with the increment of $\lambda$. This is due to the fact that in NSs, $-p'$ diminishes with the increment of $\lambda$, thus indicating that the pressure decays slower with the increment of the radial coordinate, thereby obtaining a larger total radius. On the other hand, in SSs the increment of $\lambda$ yields a larger $-p'$, indicating that the pressure decays faster with the growth of the radial coordinate, thus yielding a lower total radius.

\section{Discussion and conclusions}\label{sec:dis}

The equilibrium static configuration of NSs and SSs in the $f(R,T)$ gravity was here studied. It was considered that 
the fluid pressure is computed by $p=\omega\rho^{5/3}$ and $p=0.28(\rho-4{\cal B})$ for the NS and SS cases, 
respectively. We have started by deriving the TOV equation for the $f(R,T)$ theory of gravity. This hydrostatic 
equilibrium equation is a generalization of the one found in GR, since it presents extra terms coming from 
$2\lambda T$, considered for the functional form of the function $f(R,T)$. The equilibrium configurations of NSs 
and SSs were analyzed for different values of $\lambda$ and central energy densities.

In the modified theory of gravity under study, for a range of central energy densities, higher values of $\lambda$  yield NSs with larger mass and radius, and SSs with larger mass and lower radius. This can be observed in Figs.\ref{RhoxM_lam}-\ref{rho_R_lam}. 

We found that for the maximum value of $\lambda$ used, the maximum masses of NSs and SSs are respectively $1.538\,M_{\odot}$ and $2.017\,M_{\odot}$. These maximum mass values are determined in central energy densities $\sim30\%$ lower than those used to found the maximum mass values in the case $\lambda=0$. 

We can also observe that the values of $\lambda$ used for NSs are the tenth of that used for SSs. This indicates that in NSs, the value of $\lambda$ does not need to be large to obtain considerable effects in their structures.

Furthermore, for a more adequate form of the EoS describing NSs and for a certain value of $\lambda$, the present model is able to explain the massive pulsars PSR J$1614-2230$ and PSR J$0348+0432$ recently observed \cite{demorest_nature,antoniadis_science}. Thus, it is worthwhile to mention that only certain combinations of the EoS and $\lambda$ can be ruled out if the maximum mass found is below the observed pulsar masses.

We quote here that recently a similar approach to SSs was presented in \cite{astashenok/2015} for $f(R)=R+\alpha R^{2}$ 
gravity, with $\alpha$ being a constant. In a sense, the outcomes obtained as a consequence of the extra material term 
$2\lambda T$ presented in this article may be compared with those obtained from the consideration of the geometrical 
extra term $\alpha R^{2}$ in \cite{astashenok/2015}. In the scope of this comparison, it is worth mentioning that the 
constants of the MIT bag model have, purposely, the same values in both works. 

We have shown that by increasing the value of $\lambda$, one increases the contribution coming from the extra material 
term and obtains greater values for the masses of the SSs. In \cite{astashenok/2015}, for the $f(R)$ gravity, with $f(R)=R+\alpha R^{2}$, it was shown that the mass of a SS also increases if one considers greater values of the free parameter $\alpha$, i.e., greater contributions from the extra geometrical term. In this sense, it would not be straightforward to distinguish SS equilibrium configuration outcomes resulting from extra material or geometrical terms.

However, in the $f(R)$ gravity case, the star configuration with maximal mass corresponds to larger central densities 
when compared to GR. On the other hand, in $f(R,T)$ gravity, those values are below the GR outcomes, making possible 
to distinguish the $f(R,T)$ gravity predictions from those obtained via $f(R)$ formalism.

\

{\bf Acknowledgements}
PHRSM thanks S\~ao Paulo Research Foundation (FAPESP), grant 2015/08476-0, for financial support. JDVA thanks Coordena\c{c}\~ao de Aperfei\c{c}oamento de Pessoal de N\'\i vel Superior - CAPES, Brazil, for a grant. MM thanks FAPESP for financial support, grant $13/26258-4$.

\end{document}